\documentclass{article}
\usepackage{LaThuileFPSpro}
\usepackage{epsfig}

\begin{document}
\title{
  RECENT RESULTS FROM NA48
  }
\author{
  Spasimir Balev        \\
  {\em Joint Institute for Nuclear Research, 141980 Dubna, Russia} \\
  }
\maketitle

\baselineskip=11.6pt

\begin{abstract}
Recent results from the experiments NA48, NA48/1 and NA48/2 are
presented, including: direct $CP$-violation and Dalitz plot slopes
measurements for $K^\pm\rightarrow3\pi$ decays; $\pi\pi$ scattering
effects in $K^\pm\rightarrow\pi^\pm\pi^0\pi^0$ and $K^{+-}_{e4}$
decays, as well as $K_{e4}^\pm$ form factors and branching fraction;
measurements on radiative charged kaons and hyperon decays,
semileptonic decays of neutral and charged kaons;
$K_L\rightarrow\pi^+\pi^-$ branching ratio and $|\eta_{+-}|$
measurement; lepton universality check with $K^\pm_{l2}$ decays.
\end{abstract}
\newpage
\section{Introduction}

The series of experiments NA48, having a multipurpose large samples of
neutral and charged kaon decays, continues to provide new results in
the field of Kaon physics. In this paper briefly are described some
of the recent measurements from all three stages of the experimental
program: NA48, NA48/1 and NA48/2. The experiment NA48 (1997-2000) was dedicated to the measurement of direct $CP$-violation in $K^0$ decays. The next stage, NA48/1 (2002), was orientated mainly to the study of rare $K_S$ decays. The final
stage, NA48/2 (2003-2004), was designed to search for direct
$CP$-violation in $K^\pm$ decays (see section~\ref{ag}). Besides
these central topics, many other analyses were performed.

\section{Experimental setup}
\label{aparatus}

The NA48 beam line was designed to produce and transport both $K_L$
and $K_S$ beams simultaneously. A description of the beam line, as
well as of the NA48 detectors, can be found in\cite{Fanti:2007vi}.
Two of the measurements presented in this paper are performed during a
dedicated 1999 NA48 run. The $K_L$ beam was produced by SPS 450
GeV/$c$ proton beam on a beryllium target. The beginning of the
decay volume was defined by the last of three collimators, located
126 m downstream of the target.

For the NA48/1 experiment the $K_L$ beam was removed and the proton
flux on the $K_S$ target was greatly increased. A 24 mm platinum
absorber was placed after the Be target to reduce the photon flux in
the neutral beam. A beam of long-lived neutral particles ($\gamma$,
$n$, $K^0$, $\Lambda$ and $\Xi^0$) was selected by the sweeping
magnet, installed across the 5.2 m long collimator.

\begin{figure}[tb]
\vspace{-6mm}
\begin{center}
{\resizebox*{\textwidth}{!}{\includegraphics{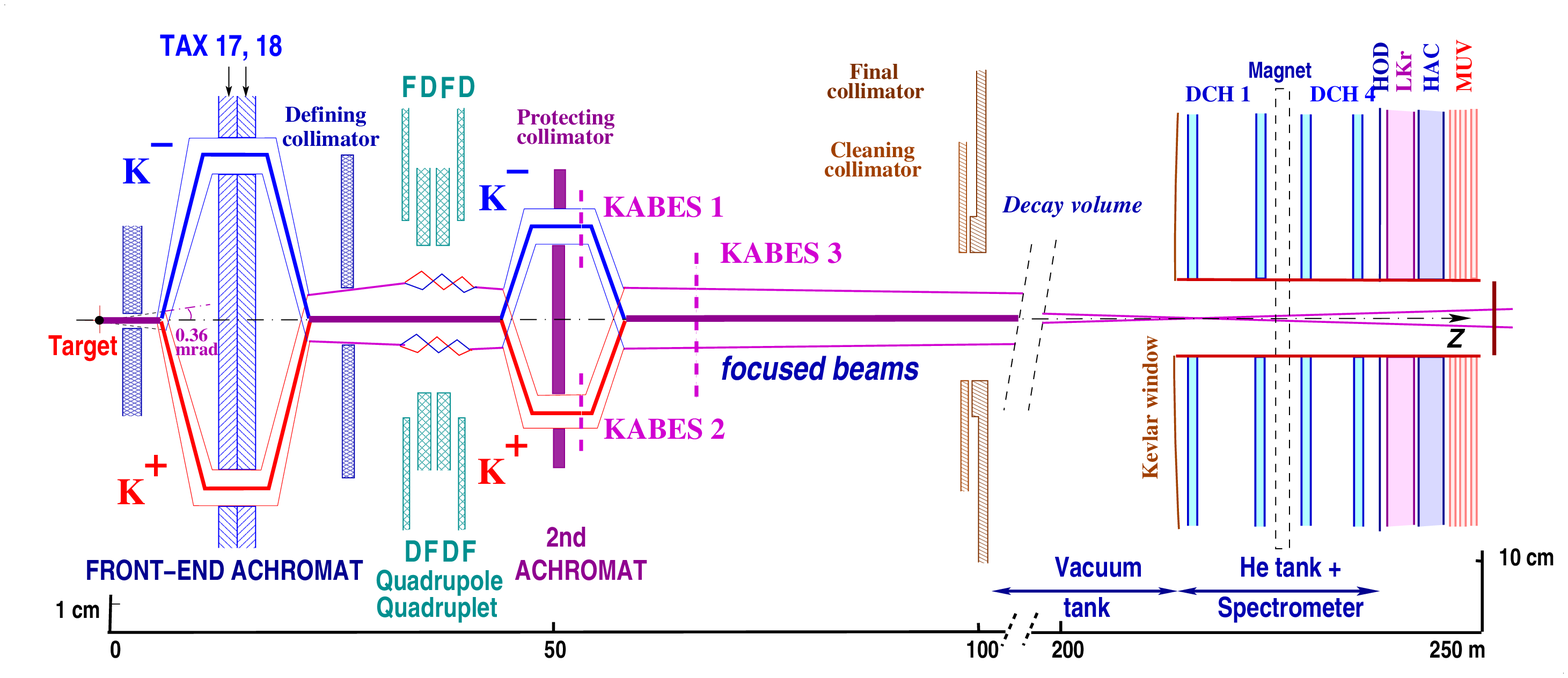}}}
\end{center}
\vspace{-6mm} \caption{\it{Schematic side view of the NA48/2 beam
line (TAX17,18: motorized beam dump/collimators used to select the
momentum of the $K^+$ and $K^-$ beams; FDFD/DFDF: focusing
quadrupoles, KABES1--3: kaon beam spectrometer stations), decay
volume and detector (DCH1--4: drift chambers, HOD: hodoscope, LKr:
EM calorimeter, HAC: hadron calorimeter, MUV: muon veto). Thick
lines indicate beam axes, narrow ones the projection of their
envelopes. Note that the vertical scales are different in the two
parts of the figure.}} \label{beams}
\end{figure}

The neutral beams were replaced by simultaneous $K^+$ and $K^-$ beams for the NA48/2 experiment. The momentum $(60\pm3)$ GeV/$c$ was formed symmetrically for $K^+$ and $K^-$ in the first achromat (see Fig.~\ref{beams}), in which the two beams
were split in the vertical plane. In the second achromat were placed
two of the three stations of the Kaon beam spectrometer (KABES). The beams followed the same path in the decay volume, comprised in a 114 m long cylindrical vacuum tank. The beam axes coincided to 1 mm, while their lateral size is about 1 cm.

The NA48 detectors, used in the presented analyses, are:
\begin{itemize}
\item a magnetic spectrometer for charged particles reconstruction,
with 4 drift chambers; the momentum resolution is
$\sigma_p/p=(1.02\oplus0.044p)\%$, where $p$ is in GeV/$c$;
\item a charged hodoscope, with good time resolution, which sends
fast trigger signals;
\item calorimeter with an active volume of 10 m$^3$ liquid krypton (LKr)
with energy resolution of
$\sigma_E/E=0.032/\sqrt{E}\oplus0.09/E\oplus0.0042$ and space
resolution of $\sigma_x=\sigma_y=0.42/\sqrt{E}\oplus0.06$ cm, where
the energy $E$ is in GeV;
\item a muon detector.
\end{itemize}

\section{Search for direct $CP$-violation in $K^\pm\rightarrow3\pi$
decays} \label{ag}

One of the most promising observables for direct $CP$-violation in
Kaon physics is the asymmetry between $K^+$ and $K^-$ decaying to three
pions. Usually, the matrix element of $K^\pm\rightarrow3\pi$ decays
is parameterized in the following form:
\begin{equation}
|M(u,v)|^2\sim1+gu+hu^2+kv^2+..., \label{eq:mel}
\end{equation}
where $g$, $h$ and $k$ are the slope parameters. The
Dalitz-variables are defined as $u=(s_3-s_0)/m_\pi^2$ and
$v=(s_1-s_2)/m_\pi^2$, where $m_\pi$ is the charged pion mass,
$s_i=(p_K-p_i)^2$, $s_0=\sum s_i/3$ ($i=1,2,3$), $p_K$ and $p_i$ are
kaon and $i$-th pion four-momenta respectively. The index $i=3$
corresponds to the odd pion, i.e. the pion with a charge different
from the other two. The parameter of direct $CP$-violation is
usually defined as
$$
A_g=\frac{g^+-g^-}{g^++g^-},
$$
where $g^+$ and $g^-$ are the linear coefficients in (\ref{eq:mel})
for $K^+$ and $K^-$ respectively. The experimental precision for
such asymmetry for both modes, $K^\pm\rightarrow\pi^\pm\pi^+\pi^-$
and $K^\pm\rightarrow\pi^\pm\pi^0\pi^0$, is at the level of
$10^{-3}$. The Standard Model (SM) predictions are below few
$10^{-5}$~\cite{5}, however some theoretical calculations involving processes
beyond the SM do not exclude enhancements of the asymmetry~\cite{7}. The main
goal of NA48/2 experiment was a search for direct $CP$-violation at
the level of $2\cdot10^{-4}$ in both $3\pi$ decay modes.

The method of such a high precision asymmetry measurement is based
on direct comparison of the $u$-ratios for $K^+$ and $K^-$ decays in
which the main possible systematic effects cancel due to the
presence of simultaneous $K^+$ and $K^-$ beams and the frequent
alternation of the magnet polarities in the beam optics and in the
magnetic spectrometer. In the $K^\pm\rightarrow\pi^\pm\pi^+\pi^-$
selection, only the magnetic spectrometer was involved in the
reconstruction of the events, while the analysis in
$K^\pm\rightarrow\pi^\pm\pi^0\pi^0$ mode was based mainly on the
information from a charge blinded detector --- LKr.

In total $\sim3.1$ billion $K^\pm\rightarrow\pi^\pm\pi^+\pi^-$ and
$\sim91$ million $K^\pm\rightarrow\pi^\pm\pi^0\pi^0$ decays were
collected during 2003 and 2004 runs and the final result on
asymmetries $A_g^c$ and $A_g^n$, respectively, yields:
$$
A_g^c=(-1.5\pm1.5_{stat}\pm0.9_{trig}\pm1.1_{syst})\cdot10^{-4},
$$
$$
A_g^n=(1.8\pm1.7_{stat}\pm0.5_{syst})\cdot10^{-4}.
$$

Both measurements are limited by the statistics and are one order of
magnitude more accurate than the previous experiments. The observed
results are compatible with the SM predictions. The method of measurement, selection of the events and the studies of main systematic contributions are described in more details in\cite{Batley:2006mu}
and\cite{Batley:2006tt}.

\section{Dalitz plot slopes measurement in
$K^\pm\rightarrow\pi^\pm\pi^+\pi^-$}

The last measurements of the Dalitz slopes $g$, $h$ and $k$ in
(\ref{eq:mel}) for $K^\pm\rightarrow\pi^\pm\pi^+\pi^-$ decay mode
are 30 years old. NA48/2 performed a new high precision measurement
in order to verify the validity of the parameterization
(\ref{eq:mel}).

Approximately $4.7\cdot10^8$ $K^\pm\rightarrow\pi^\pm\pi^+\pi^-$
decays were selected for the analysis. The measurement method is
based on fitting of the binned reconstructed $(u,|v|)$ data
distribution with a sum of four MC components generated according to
the four terms in the polynomial expansion. The free parameters in
the fitting procedure are the slopes $g$, $h$ and $k$, and the
overall normalization parameter.

The obtained results, ignoring radiative effects (apart from Coulomb
factor) and strong rescaterring effects, are:
$$
g=(-21.134\pm0.014)\%,~~h=(1.848\pm0.039)\%,~~k=(-0.463\pm0.012)\%.
$$
The values are with precision one order of magnitude better, than
the previous experiments and are in agreement with the world
averages. This is the first measurement of non-zero value of $h$.
The quality of the fit of $(u,|v|$) distribution ($\chi^2/n.d.f.=1669/1585$, yielding a satisfactory probability of $7.0\%$) shows that the polynomial
parameterization (\ref{eq:mel}) is still acceptable at an improved
level of precision (the rescaterring effects are much weaker than in
$K^\pm\rightarrow\pi^\pm\pi^0\pi^0$ mode). No significant higher
order slope parameters were found. More information on this analysis
can be found in\cite{Batley:2007md}.

\section{$\pi\pi$ scattering effects}

The single-flavour quark condensate $\langle0|\bar{q}q|0\rangle$ is
a fundamental parameter of $\chi$PT, which determines the relative
size of mass and momentum terms in the perturbative expansion. It is
a free parameter in the theory and must be determined
experimentally. The relation between $\langle0|\bar{q}q|0\rangle$
and $S$-wave $\pi\pi$ scattering lengths $a_0^0$ and $a_0^2$ in
isospin states $I=0$ and $I=2$, correspondingly, are known with
precision of $\sim2\%$\cite{Colangelo:2005cq}, so the experimental
measurement of $a_0^0$ and $a_0^2$ provides an important constraints
for $\chi$PT Lagrangian parameters. In the framework of NA48/2, the
scattering lengths can be measured from
$K^\pm\rightarrow\pi^\pm\pi^0\pi^0$ and $K^\pm\rightarrow\pi^+\pi^-
e\nu$ ($K_{e4}^{+-}$) decays.

\subsection{Rescattering effects in
$K^\pm\rightarrow\pi^\pm\pi^0\pi^0$}

\begin{figure}[t]
\begin{tabular}{cc}
{\resizebox*{0.48\textwidth}{!}{\includegraphics{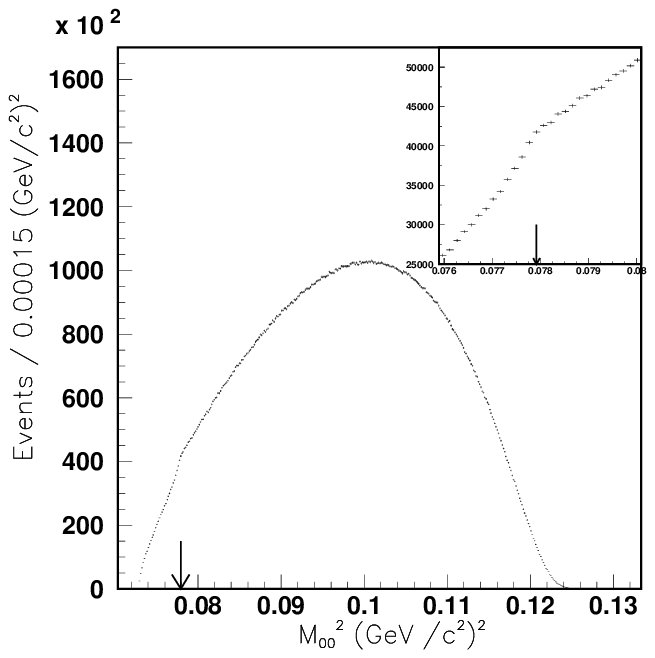}}} &
{\resizebox*{0.48\textwidth}{!}{\includegraphics{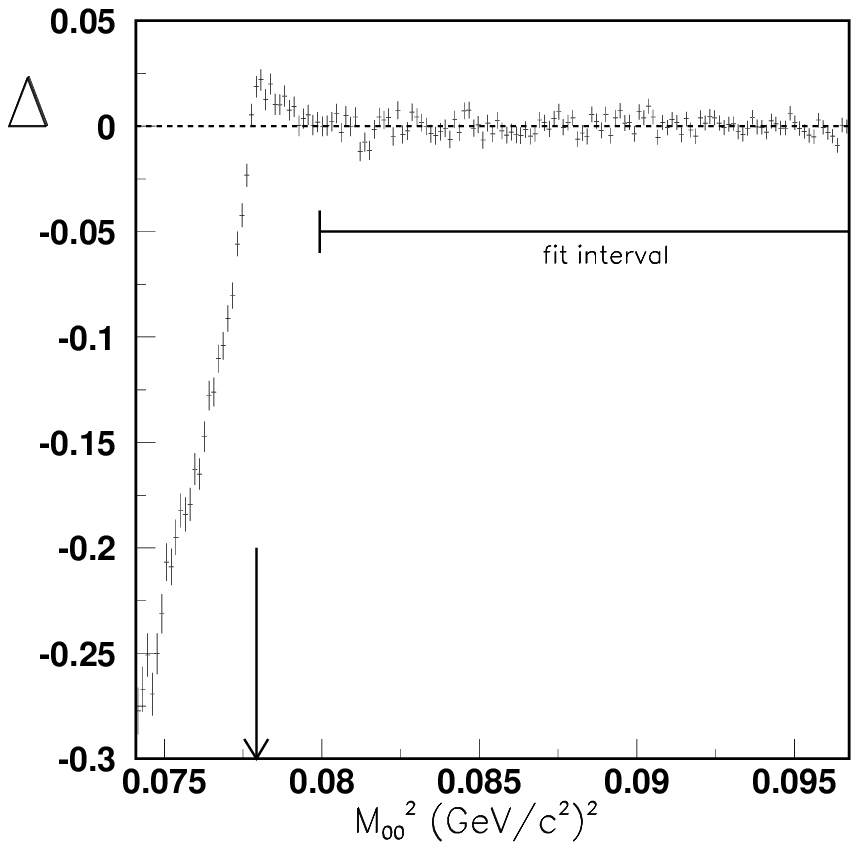}}}\\
\end{tabular}
\put(-260,-90){{\it a}}
\put(-80,-90){{\it b}}
\caption{\it{The observed "cusp"-like effect, a, and the fit result
by using the parameterization (\ref{eq:mel}), b.}} \label{fig:cusp}
\end{figure}

During the analysis of $\sim23$ million
$K^\pm\rightarrow\pi^\pm\pi^0\pi^0$ decays, taken in 2003, a sudden
"cusp" like structure was found in the spectra of $\pi^0\pi^0$
invariant mass at $M^2_{00}=4m_{\pi^\pm}^2$ (see
fig.~\ref{fig:cusp}, {\it a}). A fit to the data with the
parameterization (\ref{eq:mel}) (see fig.~\ref{fig:cusp}, {\it b})
yielded an unacceptable probability of $\chi^2/n.d.f.=9225/149$, while
the area above the "cusp" was well described
($\chi^2/n.d.f.=133/110$).

A model at one loop was developed\cite{Cabibbo:2004gq} in order to
explain the effect. The "cusp" effect is explained as a result of destructive
interference below the threshold between the two amplitudes: the
direct emission amplitude $M_0\sim1+gu/2+h'u^2/2+k'v^2/2$ and the
amplitude $M_1$, which describes the charge exchange
$\pi^+\pi^-\rightarrow\pi^0\pi^0$ in final state of
$K^\pm\rightarrow\pi^\pm\pi^+\pi^-$. A more complete formulation of
the model, which includes all rescattering processes at one loop and
two loop levels with precision $\sim5\%$\cite{Cabibbo:2005ez}, has
been used to extract the NA48/2 result:
$$
g=0.645\pm0.004_{stat}\pm0.009_{syst}
$$
$$
h'=-0.047\pm0.012_{stat}\pm0.011_{syst}
$$
$$
a_0^2=-0.041\pm0.022_{stat}\pm0.014_{syst}
$$
$$
a_0^0-a_0^2=0.268\pm0.010_{stat}\pm0.004_{syst}\pm0.013_{ext},
$$
assuming $k'=0$. The values for the scattering lengths are in good
agreement with the
theory\cite{Colangelo:2000jc}\cite{Pelaez:2004vs}. More details
about this analysis can be found in\cite{Batley:2005ax}.

Currently, there are two measurements of $k$ in (\ref{eq:mel}),
which contradicts to each other\cite{PDG}. Taking $k'$ as a free
parameter in the fit far from "cusp", a nonzero preliminary value
was obtained
$$
k'=0.0097\pm0.0003_{stat}\pm0.0008_{syst}
$$
and no change of $a_0^0$ and $a_0^2$ was observed.

\subsection{$K_{e4}$ decays form-factors}

\begin{figure}[t]
\begin{tabular}{ll}
{\resizebox*{0.48\textwidth}{!}{\includegraphics{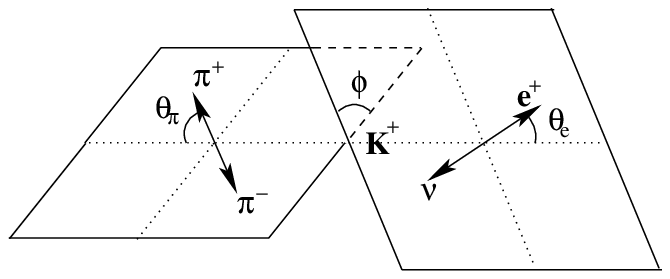}}} &
{\resizebox*{0.48\textwidth}{!}{\includegraphics{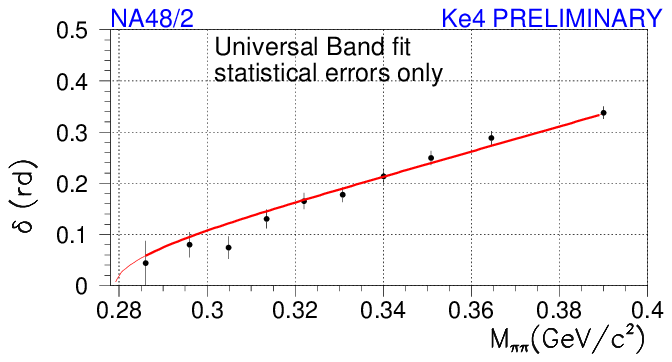}}}\\
\end{tabular}
\put(-260,-50){{\it a}}
\put(-80,-50){{\it b}}
\caption{\it{a) Definition of the angle kinematic variables,
describing $K_{e4}$ decays; b) $\delta(M_{\pi\pi})$ distribution,
fitted to obtain $a_0^0$.}} \label{ke4}
\end{figure}

The form factors of $K_{e4}^{+-}$ decay can be parameterized as a
function of five kinematic variables: the invariant masses
$M_{\pi\pi}$ and $M_{e\nu}$, and the angles $\theta_\pi$, $\theta_e$
and $\phi$ (see Fig.~\ref{ke4}). The hadronic part in the matrix
element can be described in terms of two axial ($F$ and $G$) and one
vector ($H$) form factors\cite{dafne}. Their expansion into partial
$s$ and $p$ waves (neglecting $d$ waves) and into a Taylor series in
$q^2=M^2_{\pi\pi}/4m^2_{\pi^\pm}-1$ allows a measurement of the form
factor parameters from the experimental data\cite{ke4a}\cite{ke4b}:
$$
F=F_se^{i\delta_s}+F_p\cos\theta_\pi
e^{i\delta_p},~~G=G_pe^{i\delta_g},~~H=H_pe^{i\delta_h},
$$
where
$$
F_s=f_s+f'_sq^2+f''_sq^4,~~F_p=f_p+f'_pq^2+...,
$$
$$
G_p=g_p+g'_pq^2+...,~~H_p=h_pe+h'_pq^2+...
$$

Analysing part of 2003 data, $3.7\cdot10^5$ $K^{+-}_{e4}$ decays
were selected with background of $0.5\%$, mainly from $3\pi$ decays
with $\pi\rightarrow e\mu$ or pion mis-identification.

The following method was used to extract the form factor parameters:
In a first step, in
($M_{\pi\pi}$,$M_{e\nu}$,$\cos\theta_\pi$,$\cos\theta_e$,$\phi$)
space were defined 10x5x5x5x12 iso-populated bins. For each bin in
$M_{\pi\pi}$, comparing data and MC, ten independent five-parameter
($F_s,F_p,G_p,H_p,\delta=\delta_s-\delta_p$) fits were performed. In
the second step a fit of the distributions in $M_{\pi\pi}$ was
performed to extract the form factor parameters. The
$\delta(M_{\pi\pi})$ distribution was fitted with a one-parameter
function given by the numerical solution of the Roy
equations\cite{roy}, in order to determine $a^0_0$, while $a^2_0$
was constrained to lie on the centre of the universal band. The
following preliminary results were obtained:
$$
f'_s/f_s = 0.169\pm0.009_{stat}\pm0.034_{syst}
$$
$$
f''_s/f_s =-0.091\pm0.009_{stat}\pm0.031_{syst}
$$
$$
f_p/f_s =-0.047\pm0.006_{stat}\pm0.008_{syst}
$$
$$
g_p/f_s = 0.891\pm0.019_{stat}\pm0.020_{syst}
$$
$$
g'_p/f_s = 0.111\pm0.031_{stat}\pm0.032_{syst}
$$
$$
h_p/f_s =-0.411\pm0.027_{stat}\pm0.038_{syst}
$$
$$
a^0_0 = 0.256\pm0.008_{stat}\pm0.007_{syst}\pm0.018_{theor},
$$
where the systematic uncertainty was determined by comparing two
independent analyses and taking into account the effect of the
reconstruction method, acceptance, fit method, uncertainty on
background estimate, electron identification efficiency, radiative
corrections and bias due to the neglected $M_{e\nu}$ dependence. The
form factors are measured relative to $f_s$, which is related to the
decay rate. The obtained value for $a^0_0$ is compatible with the
$\chi$PT prediction\cite{Colangelo:2001df} and with previous
experiments\cite{Pislak:2003sv}.

The form factors were measured also for
$K^\pm\rightarrow\pi^0\pi^0e\nu$ ($K^{00}_{e4}$) decays, on $\sim 10^4$ selected events from 2003 run and $\sim3\cdot10^4$
events from 2004 run, using the same formalism. Due to symmetry of
$\pi^0\pi^0$ system, the $P$-wave is not present and only two
parameter are left:
$$
f'_s/f_s=0.129\pm0.036_{stat}\pm0.020_{syst}
$$
$$
f''_s/f_s=-0.040\pm0.034_{stat}\pm0.020_{syst}.
$$
The preliminary result is compatible with $K^{+-}_{e4}$.

In addition, the branching fraction of $K^{00}_{e4}$ was measured by
using only 2003 data, normalising to
$K^\pm\rightarrow\pi^\pm\pi^0\pi^0$:
$$
BR(K^{00}_{e4})=(2.587\pm0.026_{stat}\pm0.019_{syst}\pm0.029_{ext})\cdot10^{-5},
$$
where the systematic uncertainty takes into account the effect of
acceptance, trigger efficiency and energy measurement of the
calorimeter, while the external uncertainty is due to the
uncertainty on the $K^\pm\rightarrow\pi^\pm\pi^0\pi^0$ branching
fraction. The result is $\sim8$ times more precise than the previous
measurement.

\section{New measurements on kaon and hyperon radiative decays}

\subsection{$K^\pm\rightarrow\pi^\pm\pi^0\gamma$ measurements}

A measurement of Direct photon Emission (DE) with respect to Inner
Brems\-strahlung (IB) and the interference (INT) between these two
amplitudes was performed on a subsample of NA48/2 collected during
2003 run. The $K^\pm\rightarrow\pi^\pm\pi^0\gamma$ are
described in terms of two kinematic variables: the energy of charged
pion in kaon center of mass system ($T^*_\pi$), and
$W^2=(p_Kp_\gamma)(p_\pi p_\gamma)/m^2_\pi m^2_K$, where $p_K$,
$p_\pi$ and $p_\gamma$ are the four-momenta of the kaon, charged
pion and the odd gamma. About $124\cdot10^3$ events were selected in
the ranges $T^*_\pi<80$ MeV and $0.2<W<0.9$. In the previous
measurements a lower cut $T^*>55$ MeV was introduced in order to
suppress $K^\pm\rightarrow\pi^\pm\pi^0\pi^0$ and
$K^\pm\rightarrow\pi^\pm\pi^0$ background. In NA48/2 measurement
these backgrounds are avoided with an application of a special
algorithm, which detects overlapping gamma in the detector and a
maximum allowed deviation of reconstructed $K$ mass $\pm 10$ MeV
from its nominal value. The upper cut on $T^*_\pi$ rejects
$K^\pm\rightarrow\pi^\pm\pi^0$ decays. The background in the
selected sample is kept under $10^{-4}$. The photon mistagging
(i.e., choice of wrong odd photon) is estimated to be less than
0.1\%.

The preliminary results:
$$
\rm{Frac(DE)}=(3.35\pm0.35_{stat}\pm0.25_{syst})\%
$$
$$
\rm{Frac(INT)}=(-2.67\pm0.81_{stat}\pm0.73_{syst})\%
$$
are obtained by using extended maximum likelihood method: the
experimental $W$ distribution is fitted with proportionally
simulated IB, DE and INT distributions. The systematic error is
dominated by the trigger efficiency.

This is the first observation of non zero INT component.

\subsection{First observation of $K^\pm\rightarrow\pi^\pm\gamma
e^+e^-$}

\begin{figure}[tb]
\vspace{-6mm}
\begin{center}
{\resizebox*{0.7\textwidth}{!}{\includegraphics{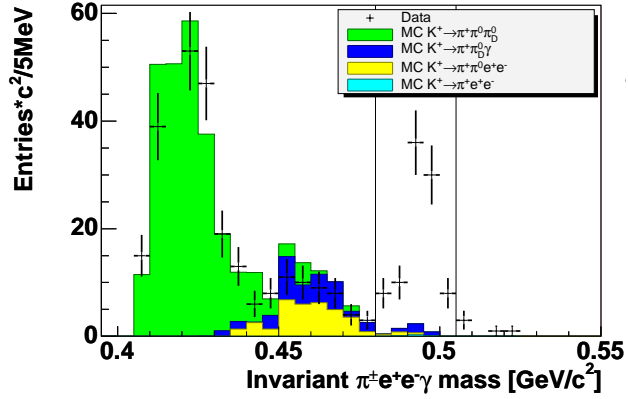}}}
\end{center}
\vspace{-6mm} \caption{\it{The invariant mass of $\pi^\pm
e^+e^-\gamma$, together with the simulated background.}}
\label{pgee}
\end{figure}

NA48/2 experiment observed for the first time the radiative decay
$K^\pm\rightarrow\pi^\pm\gamma e^+e^-$. 92 candidates were selected,
with $1\pm1$ accidental background and $5.1\pm1.7$ misidentification
background (Fig.~\ref{pgee}). By using
$K^\pm\rightarrow\pi^\pm\pi^0$ as normalization channel the
branching ratio preliminary was estimated to be
$$
BR(K^\pm\rightarrow\pi^\pm\gamma
e^+e^-)=(1.27\pm0.14_{stat}\pm0.05_{syst})\cdot10^{-8}.
$$

\subsection{First observation of $\Xi^0\rightarrow\Lambda^0e^+e^-$}

\begin{figure}[tb]
\vspace{-6mm}
\begin{center}
{\resizebox*{0.6\textwidth}{!}{\includegraphics{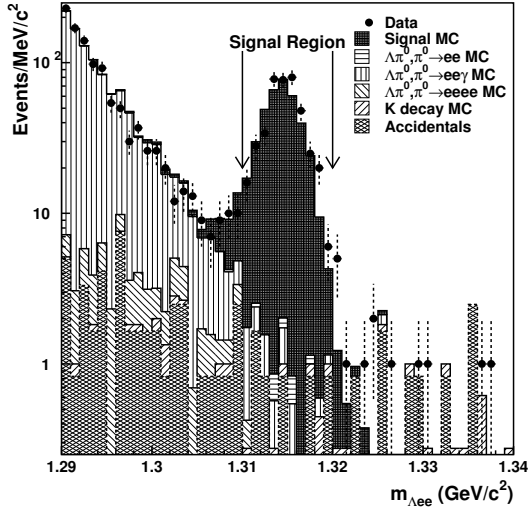}}}
\end{center}
\vspace{-6mm} \caption{\it{The invariant mass of $\Lambda e^+e^-$,
together with the simulated background.}} \label{xi2lamee}
\end{figure}

In the 2002 NA48/1 run the weak radiative decay
$\Xi^0\rightarrow\Lambda^0e^+e^-$ was detected for the first
time\cite{Batley:2007hp}. 412 candidates were selected with 15
background events (Fig.~\ref{xi2lamee}). The obtained branching
fraction
$$
BR(\Xi^0\rightarrow\Lambda^0e^+e^-)=(7.7\pm0.5_{stat}\pm0.4_{syst})\cdot10^{-6}
$$
is consistent with inner bremsstrahlung-like $e^+e^-$ production
mechanism.

The decay parameter $\alpha_{\Xi\Lambda ee}$ can be measured from
the angular distribution
$$
dN/d\cos\theta_{p\Xi}=\frac{N}{2}(1-\alpha_{\Xi\Lambda
ee}\alpha_-\cos\theta_{p\Xi}),
$$
where $\theta_{p\Xi}$ is the angle between the proton from
$\Lambda\rightarrow p\pi$ decay relative to the $\Xi^0$ line of
flight in the $\Lambda$ rest frame. The obtained value
$$
\alpha_{\Xi\Lambda ee} = -0.8\pm0.2
$$
is consistent with the latest published value of the decay asymmetry
parameter for $\Xi\rightarrow\Lambda\gamma$.

\section{New measurements of $K_L$ decays}

In 1999 a dedicated NA48 run employed a minimum bias trigger to
collect semileptonic decays of $K_L$. Two new measurements from this
run are presented.

\subsection{$K_{L\mu3}$ form factors}

$K_{l3}$ decays provide the cleanest way to extract $|V_{us}|$
element in the CKM matrix. Recent calculations in the framework of
$\chi$PT show how the vector form factor at zero momentum transfer,
$f_+(0)$, can be constrained experimentally from the slope and
curvature of the scalar form factor $f_0$ of the $K_{\mu3}$ decay.
In addition, these form factors are needed to calculate the phase
space integrals, which are used in $|V_{us}|$ determination.

\begin{figure}[tb]
\vspace{-6mm}
\begin{center}
{\resizebox*{0.6\textwidth}{!}{\includegraphics{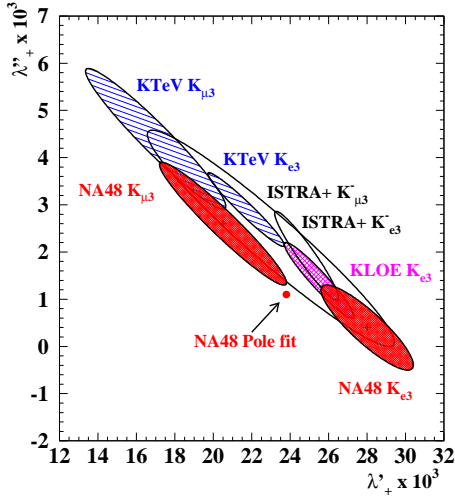}}}
\end{center}
\vspace{-6mm} \caption{\it{Comparison between recent results on
$K_{Ll3}$ form factors measurements.}} \label{kmu3}
\end{figure}

Approximately $2.6\cdot10^6$ $K_{\mu3}$ decays were selected from
the 1999 minimum bias run. By studying the Dalitz plot density, the
following slopes for the vector and the scalar form factors were
obtained
$$
\lambda'_+=(20.5\pm2.2_{stat}\pm2.4_{syst})\cdot10^{-3}
$$
$$
\lambda''_+=(2.6\pm0.9_{stat}\pm1.0_{syst})\cdot10^{-3}
$$
$$
\lambda_0=(9.5\pm1.1_{stat}\pm0.8_{syst}.
$$
The results show a presence of quadratic term in the expansion of
the vector form factor in agreement with other recent measurements.
A comparison between the results of the quadratic fits as reported
by the recent experiments is presented in Fig.~\ref{kmu3}.

The results obtained with linear fit are
$$
\lambda_+=(26.7\pm0.6_{stat}\pm0.8_{syst})\cdot10^{-3}
$$
$$
\lambda_0=(11.7\pm0.7_{stat}\pm1.0_{syst})\cdot10^{-3}.
$$
The value for $\lambda_+$ is well compatible with the recent KTeV
measurement, while $\lambda_0$ is shifted towards lower values.
Details on NA48 $K_{\mu3}$ measurement can be found
in\cite{Veltri:2007jm}.

\subsection{$\Gamma(K_L\rightarrow\pi^+\pi^-)/\Gamma(K_{Le3})$ ratio and
$|\eta_{+-}|$ measurements}

The recent results on $\Gamma(K_L\rightarrow\pi^+\pi^-)$ and the $CP$-violation parameter $|\eta_{+-}|$, performed by the experiments KTeV and
KLOE, and the measurement of the ratio $\Gamma(K_L\rightarrow\pi^+\pi^-)/\Gamma(K_{Le3})$ by KTeV disagree with 2004 edition of PDG\cite{pdg4} by 10\% and 5\%
respectively (or more than four standard deviations). Additional
information could clarify the situation.

During the dedicated 1999 NA48 run $\sim47\cdot10^3$
$K_L\rightarrow\pi^+\pi^-$ and $\sim5\cdot10^6$ $K_{Le3}$ decays
were collected. The ratio
$\Gamma(K_L\rightarrow\pi^+\pi^-)/\Gamma(K_{Le3})$ is measured to be
$$
\frac{\Gamma(K_L\rightarrow\pi^+\pi^-)}{\Gamma(K_{Le3})}=(4.835\pm0.022_{stat}\pm0.016_{syst})\cdot10^{-3}.
$$
For $BR(K_L\rightarrow\pi^+\pi^-)$ and $|\eta_{+-}|$ calculation the $CP$-conserving 
direct emission $K_L\rightarrow\pi^+\pi^-\gamma$ contribution to the
$K_{2\pi}$ signal was estimated and subtracted. The branching
fraction of $K_{2\pi}$, including only the inner bremsstrahlung
radiative component was measured to be
$$
BR(K_L\rightarrow\pi^+\pi^-+\pi^+\pi^-\gamma(IB))=(1.941\pm0.019)\cdot10^{-3}.
$$
Using this result and the most precise single measurements of
$\tau_{K_S}$ (by NA48), $\tau_{K_L}$ and
$BR(K_S\rightarrow\pi^+\pi^-)$ (by KLOE), the $CP$-violation parameter
$|\eta_{+-}|$ is calculated:
$$
|\eta_{+-}|=\sqrt\frac{\tau_{K_S}BR(K_L\rightarrow\pi^+\pi^-)}{\tau_{K_L}BR(K_S\rightarrow\pi^+\pi^-)}=(2.223\pm0.012)\cdot10^{-3}.
$$
All the presented results are in agreement with the recent KTeV and
KLOE results. Details on the analysis can be found
in\cite{Lai:2006cf}.

\section{Results from $K^{+-}$ semileptonic decays}

The branching ratios of semileptonic kaon decays are needed to
determine $|V_{us}|$ element in the CKM matrix. In addition
$\Gamma(K_{e3})/\Gamma(K_{\mu3})$ is a function of the slope
parameters of the form factors, which can be used for consistency
check under the assumption of $\mu-e$ universality.

During 2003 data taking of NA48/2 a special run was dedicated to collect
semileptonic decays. Approximately 56000 $K^+_{e3}$, 31000
$K^-_{e3}$, 49000 $K^+_{\mu3}$, 28000 $K^-_{\mu3}$, 462000
$K^+_{2\pi}$ and 256000 $K^-_{2\pi}$ decays were selected for the
measurement. The ratios of decay widths, combined for $K^+$ and
$K^-$, are:
$$
\Gamma(K_{e3})/\Gamma(K_{2\pi})=0.2496\pm0.0009_{stat}\pm0.0004_{syst}
$$
$$
\Gamma(K_{\mu3})/\Gamma(K_{2\pi})=0.1637\pm0.0006_{stat}\pm0.0003_{syst}
$$
$$
\Gamma(K_{\mu3})/\Gamma(K_{e3})=0.656\pm0.003_{stat}\pm0.001_{syst}
$$
Taking the PDG value for the $K_{2\pi}$ branching fraction,
$0.2092\pm0.0012$, the branching fractions for the semileptonic
decays are found to be:
$$
BR(K_{e3})=(5.221\pm0.019_{stat}\pm0.008_{syst}\pm0.030_{norm})\%
$$
$$
BR(K_{\mu3})=(3.425\pm0.013_{stat}\pm0.006_{syst}\pm0.020_{norm})\%
$$
The uncertainty is dominated by the existing data for the
$BR(K_{2\pi})$. The branching fractions are higher than PDG values
for both $K_{e3}$ and $K_{\mu3}$, confirming the $K_{e3}$ results
reported by the BNL-E865 collaboration.

By using the measured values for the vector and the scalar form
factors\cite{PDG}, and assuming $e-\mu$ universality, a value
$0.6682\pm0.0017$ for the ratio $\Gamma(K_{\mu3})/\Gamma(K_{e3})$
can be estimated. The NA48/2 result suggests a lower value for
$\lambda_0$ than the current world average for $K^\pm$, as found in
recent measurements from $K_L$ decays.

The product $|V_{us}|f_+(0)$ can be calculated by using both
$K_{e3}$ and $K_{\mu3}$ measured branching ratios:

$$
\rm{From~K_{e3}:}~~|V_{us}|f_+(0)=0.2204\pm0.0012
$$
$$
\rm{From~K_{\mu3}:}~~|V_{us}|f_+(0)=0.2177\pm0.0013
$$
The errors are dominated by the uncertainties of the external
quantities needed for the calculation. Combining the results,
assuming lepton universality and taking the value of $f_+(0)$ for
neutral kaons, the obtained $|V_{us}|$ element is
$$
|V_{us}|=0.2289\pm0.0023,
$$
which is consistent with CKM matrix unitarity predictions. For detailed description of the analysis see\cite{Batley:2006cj}.

\section{$\Gamma(K^\pm_{e2})/\Gamma(K^\pm_{\mu2})$ measurement}

The ratio $R_K=\Gamma(K^\pm_{e2})/\Gamma(K^\pm_{\mu2})$ is a test
for lepton universality and $V-A$ coupling. The SM prediction is
$R_K=(2.472\pm0.001)\cdot10^{-5}$, while the current PDG average is
$R_K=(2.45\pm0.11)\cdot10^{-5}$. Recently a new important physical
motivation for a precise measurement of this ratio was
added\cite{masiero}: SUSY lepton flavour violating contributions
could shift $R_K$ by a relative amount of 2-3\%.

The NA48/2 analysis exploits the similarity between both decays to
cancel most of the possible systematic effects. In 2003 run 5239
$K_{e2}$ were selected with $\sim14\%$ background mainly from
$K_{\mu2}$. The obtained preliminary result is
$$
R_K=(2.416\pm0.043_{stat}\pm0.024_{syst})\cdot10^{-5}.
$$
The estimations yield that the combined 2003 and 2004 result will
not be sufficient to obtain a total error smaller than 1\%. A
dedicated 2007 run is in preparation. The conservative estimation
for the error, which will be reached in $R_K$ measurement is 0.7\%.
The experiment P326 could reach a per mill uncertainty, adding a new
item to its physics program.


\begin{thebibliography}{99}
\bibitem{Fanti:2007vi}
  V.~Fanti {\it et al},
  Nucl.\ Instrum.\ Meth.\  A {\bf 574}, 433 (2007).

  A. Lai {\it et al}, Eur. Phys. J. C {\bf 22} 231 (2001).

  J.R. Batley {\it et al}, Phys. Lett. B {\bf 544} 97 (2002).

\bibitem{5}
L. Maiani and N. Paver, {\it The second DA$\Phi$NE Physics Handbook,
INFN, LNF}, {\bf Vol 1}, 51 (1995).\\
E.P. Shabalin, {\it Phys. Atom. Nucl.} {\bf 68}, 88 (2005).\\
A.A. Belkov, A.V. Lanyov and G. Bohm, {\it Czech. J. Phys.} {\bf 55
Suppl. B}, 193 (2004).\\
G. D'Ambrosio and G. Isidori, {\it Int. J. Mod. Phys.} {\bf A13}, 1 (1998).\\
I. Scimemi, E. Gamiz and J. Prades, hep-ph/0405204.\\
G. F\"{a}ldt and E.P. Shabalin, {\it Phys. Lett.} {\bf B635}, 295
(2006).

\bibitem{7}
G. D'Ambrosio, G. Isidori and G. Martinelli, {\it Phys. Lett.} {\bf
B480}, 164 (2000).\\
E.P. Shabalin, ITEP-8-98 (1998).

\bibitem{Batley:2006mu}
  J.~R.~Batley {\it et al},
  Phys.\ Lett.\  B {\bf 634}, 474 (2006)
  [arXiv:hep-ex/0602014].

\bibitem{Batley:2006tt}
  J.~R.~Batley {\it et al},
  Phys.\ Lett.\  B {\bf 638}, 22 (2006)
  [Erratum-ibid.\  B {\bf 640}, 297 (2006)]
  [arXiv:hep-ex/0606007].

\bibitem{Batley:2007md}
  J.~R.~Batley {\it et al},
  arXiv:hep-ex/0702045.

\bibitem{Colangelo:2005cq}
  G.~Colangelo,
  AIP Conf.\ Proc.\  {\bf 756}, 60 (2005)
  [arXiv:hep-ph/0501107].

\bibitem{Cabibbo:2004gq}
  N.~Cabibbo,
  Phys.\ Rev.\ Lett.\  {\bf 93}, 121801 (2004)
  [arXiv:hep-ph/0405001].

\bibitem{Cabibbo:2005ez}
  N.~Cabibbo and G.~Isidori,
  JHEP {\bf 0503}, 021 (2005)
  [arXiv:hep-ph/0502130].

\bibitem{Colangelo:2000jc}
  G.~Colangelo, J.~Gasser and H.~Leutwyler,
  Phys.\ Lett.\  B {\bf 488}, 261 (2000)
  [arXiv:hep-ph/0007112].

\bibitem{Pelaez:2004vs}
  J.~R.~Pelaez and F.~J.~Yndurain,
  Phys.\ Rev.\  D {\bf 71}, 074016 (2005)
  [arXiv:hep-ph/0411334].

\bibitem{Batley:2005ax}
  J.~R.~Batley {\it et al},
  Phys.\ Lett.\  B {\bf 633}, 173 (2006)
  [arXiv:hep-ex/0511056].

\bibitem{PDG}
  W.-M. Yao {\it et al}, J. Phys. {\bf G33}, 1 (2006).

\bibitem{dafne}
  J. Bijnens {\it et al}, 2nd DA$\Phi$NE Physics Handbook, 315
  (1995).

\bibitem{ke4a}
  A. Pais and S. B. Treiman, Phys. Rev. {\bf 168}, 1858 (1968).

\bibitem{ke4b}
  G. Amoros and J. Bijnens, J. Phys. G {\bf 25}, 1607 (1999).

\bibitem{roy}
  B. Ananthanarayan {\it et al}, Phys. Rept. {\bf 353}, 207 (2001).

\bibitem{Colangelo:2001df}
  G.~Colangelo, J.~Gasser and H.~Leutwyler,
  Nucl.\ Phys.\  B {\bf 603}, 125 (2001)
  [arXiv:hep-ph/0103088].

\bibitem{Pislak:2003sv}
  S.~Pislak {\it et al.},
  Phys.\ Rev.\  D {\bf 67}, 072004 (2003)
  [arXiv:hep-ex/0301040].

\bibitem{Batley:2007hp}
  J.~R.~Batley {\it et al},
  arXiv:hep-ex/0703023.

\bibitem{Veltri:2007jm}
  M.~Veltri,
  arXiv:hep-ex/0703007.

\bibitem{pdg4}
  S. Eidelman {\it et al}, Phys. Lett. B {\bf 592} (2004) 1.

\bibitem{Lai:2006cf}
  A.~Lai {\it et al},
  Phys.\ Lett.\  B {\bf 645}, 26 (2007)
  [arXiv:hep-ex/0611052].

\bibitem{Batley:2006cj}
  J.~R.~Batley {\it et al},
  arXiv:hep-ex/0702015.

\bibitem{masiero}
  A.~Masiero, P.~Paradisi and R.~Petronzio,
  Phys.\ Rev.\ D {\bf 74} (2006) 011701
  [arXiv:hep-ph/0511289].

\end{thebibliography}
\end{document}